\documentclass[aps,prd,groupedaddress,showpacs,showkeys]{revtex4}
\usepackage{amssymb}
\usepackage{amsmath}
\usepackage{amsfonts}
\usepackage{epsfig}
\usepackage{graphicx}
\usepackage{tabularx}
\usepackage{bm}
\usepackage{mathtext}

\newcommand{\beaa}{\begin{eqnarray*}}
\newcommand{\enaa}{\end{eqnarray*}}
\newcommand{\bea}{\begin{eqnarray}}
\newcommand{\ena}{\end{eqnarray}}
\newcommand{\seq}{\begin{subequations}}
\newcommand{\sen}{\end{subequations}}
\newcommand{\eq}{\begin{eqnarray}}
\newcommand{\en}{\end{eqnarray}}

\def\shiftdown#1{#1\llap{\lower.04ex\hbox{#1}}}

\newcommand{\ra}{\rangle}
\newcommand{\la}{\langle}

\newcommand{\bfk}{{\bf k}_{\perp}}

\newcommand{\bfb}{{\bf b}_{\perp}}
\newcommand{\bfe}{{\bf e}_{\perp}}
\newcommand{\bfP}{{\bf P}_{\perp}}
\newcommand{\bfPpr}{{\bf P}_{\perp}^\prime}
\newcommand{\bfp}{{\bf p}_{\perp}}
\newcommand{\bfppr}{{\bf p}_{\perp}^\prime}

\def\arraystretch{1.5}

\begin{document}

\title{Light and heavy mesons in a soft-wall holographic approach}
\noindent
\author{
Tanja Branz$^1$,
Thomas Gutsche$^1$,
Valery E. Lyubovitskij$^1$
\footnote{On leave of absence
from Department of Physics, Tomsk State University,
634050 Tomsk, Russia},
Ivan Schmidt$^2$,
Alfredo Vega$^2$
\vspace*{1.2\baselineskip}}
\affiliation{$^1$ Institut f\"ur Theoretische Physik,
Universit\"at T\"ubingen, \\
Kepler Center for Astro and Particle Physics,
\\ Auf der Morgenstelle 14, D-72076 T\"ubingen, Germany
\vspace*{1.2\baselineskip} \\
\hspace*{-1cm}$^2$ Departamento de F\'\i sica y Centro Cient\'\i
fico Tecnol\'ogico de Valpara\'\i so (CCTVal), Universidad T\'ecnica
Federico Santa Mar\'\i a, Casilla 110-V, Valpara\'\i so, Chile
\vspace*{0.3\baselineskip}\\}

\date{\today}

\begin{abstract}

We study the mass spectrum and decay constants of light and heavy
mesons in a soft-wall holographic approach, using the correspondence
of string theory in Anti-de Sitter space and conformal field theory
in physical space-time.

\end{abstract}



\maketitle

\newpage

\section{Introduction}

In a series of papers~\cite{Brodsky:2003px}-\cite{Brodsky_ANL}
Brodsky and de T\'eramond developed a semiclassical approximation to
QCD - light-front holography (LFH)-, an approach based on the
correspondence of string theory in Anti-de Sitter (AdS) space and
conformal field theory (CFT) in physical
space-time~\cite{Maldacena:1997re,Klebanov:2009zz}. Light-front
holography~\cite{Brodsky:2003px}-\cite{Brodsky_ANL} is one of the
exciting features of the AdS/CFT correspondence. The LFH approach is
a covariant and analytic model for hadron structure with
confinement at large and conformal behavior at short
distances. It is analogous to the Schr\"odinger theory for atomic
physics. It provides a precise mapping of the string modes $\Phi(z)$ in
the AdS fifth dimension $z$ to the hadron light-front wave functions
(LFWF) in physical space-time, in terms of the light-front impact
variable $\xi$, which measures the separation of the quark and
gluonic constituents inside a hadron. Therefore, different values of
the holographic variable $z$ correspond to different scales at which
the hadron is examined. The mapping was obtained by matching certain
matrix elements (e.g. electromagnetic pion form factor, the
energy-momentum tensor) in the two approaches - string theory in AdS
and light-front theory in Minkowski space-time.

In order to break conformal invariance and incorporate confinement
in the infrared (IR) region, two alternative AdS/QCD backgrounds
have been suggested in the literature: the ``hard-wall''
approach~\cite{Polchinski:2001tt}-\cite{Grigoryan:2007vg}, based on
introducing an IR brane cutoff in the fifth dimension, and
the ``soft-wall''
approach~\cite{Brodsky:2007hb},\cite{Karch:2006pv}-\cite{Afonin:2010fr},
based on using a soft cutoff by introducing a background dilaton
field in the AdS space or using a warp factor in the metric.

Both approaches have certain advantages. One of the problems of the
``hard-wall'' scenario is a linear dependence of hadron masses with $M
\propto L$ for higher values of the orbital momentum $L$ instead of the
quadratic behavior $M^2 \propto L$ (known as Regge trajectory). In
fact, the ``soft-model'' was initiated in order to solve the problem
of the hadronic mass spectrum. In Refs.~\cite{Brodsky:2007hb} and
\cite{Karch:2006pv}-\cite{Afonin:2010fr} the ``soft-model'' has been
applied to different aspects of hadron properties, including the
hadron and glueball mass spectrum, the heavy quark potential, form
factors, deep inelastic scattering, etc. Notice that in the LFH
approach~\cite{Brodsky:2003px}-\cite{Brodsky_ANL} both scenarios for
AdS/QCD backgrounds (``hard-wall'' and ``soft-wall'') are used in
order to map the string modes to the LFWF restricting to the case of
massless quarks. A generalization of the LFWF for massive quarks has
been suggested by Brodsky and de T\'eramond, including quark masses
explicitly in the LF kinetic energy $\sum\limits_i ({\bf k}_{\perp
i}^2 + m_i^2)/x_i$ (see details in Ref.~\cite{Brodsky:2008pg}).
The LFH approach has been successfully applied to the description of the
mass spectrum of meson and baryons (reproducing the Regge
trajectories), the pion leptonic constant, the electromagnetic form factors
of pion and nucleons, etc.~\cite{Brodsky:2003px}-\cite{Brodsky_ANL}.
In Refs.~\cite{Vega:2008af,Vega:2008te} an alternative soft-wall
holographic model has been developed, which provides an extension to
hadrons with an arbitrary number of constituents. In
Ref.~\cite{Vega:2009zb} meson wave functions derived in these two
approaches were discussed in the case of massive quarks.

Notice that the LFH approach developed in
Ref.~\cite{Brodsky:2003px}-\cite{Brodsky_ANL} uses the so-called
``negative'' dilaton field profile ($e^{-\phi(z)}$ with $\phi(z) = -
\kappa^2 z^2$), which was necessary to produce a massless pion
and the behavior of the gravitational potential. In this paper we use
the soft-wall approach with a ``positive'' dilaton field profile
$\phi(z) = \kappa^2 z^2$,  as suggested originally in
Ref.~\cite{Karch:2006pv}. In the context of the original soft-wall
model~\cite{Karch:2006pv}-\cite{Afonin:2010fr}, the positive sign in
the dilaton profile is important to reproduce the correct behavior
of Regge trajectories for higher spin states. In fact, as stressed
in a recent paper~\cite{Afonin:2010fr}, the two different signs of
the dilaton field profile are related to two different ways of 
introducing higher spin fields. On the other hand, as was shown in 
Refs.~\cite{Karch:2006pv,Katz:2005ir,Kwee:2007nq,Gherghetta:2009ac},
the pion appears massless in soft-wall models with a ``positive''
dilaton. Moreover, one of the outstanding features of the LFH
approach~\cite{Brodsky:2003px}-\cite{Brodsky_ANL} is the LF mapping
-- matching of QCD LF dynamics with the corresponding string
dynamics in AdS space. This was, for example, done in the case of the
pion electromagnetic form factor (see details
in~\cite{Brodsky:2007hb}). In particular, in the calculation of the
pion form factor at large values of the Euclidean momentum squared
it was shown that the dilaton should have a positive sign.

In this manuscript we show that the use of the positive dilaton in the LFH
approach~\cite{Brodsky:2003px}-\cite{Brodsky_ANL} is possible if we
modify the mass term of the AdS$_{d+1}$ action.
The proposed approach is applied to the study of the mass spectrum
and decay properties of light and heavy mesons. In the case of the
mass spectrum we include color Coulomb and hyperfine--splitting 
corrections. The paper is structured as follows. First, in
Sec.~II, we briefly discuss the basic notions of the approach. In
Sec.~III, we consider the mass spectrum and decay properties of
light and heavy mesons. Finally, in Sec.~IV, we summarize our
results.

\section{Basic Approach}

\subsection{AdS action and Schr\"odinger equation of motion for mesons}

Our starting point is the action in AdS$_{d+1}$ spacetime for a spin-$J$ field
$\Phi_J = \Phi_{M_1 \cdots M_J}(x,z)$ -- a symmetric, traceless
tensor used in the LFH
approach~\cite{Brodsky:2003px}-\cite{Brodsky_ANL}, where we perform
two modifications: 1) we use a positive dilaton profile $\phi(z) =
\kappa^2 z^2$; 2) we include a nontrivial $z$-dependence of the mass
term coefficient $\mu^2_J \to \mu^2_J(z)$ due to the interaction of
the dilaton field with the matter field: 
\eq\label{ADS_action}
S_\Phi = \frac{(-1)^J}{2} \int d^d x dz \, \sqrt{g} \, e^{-\phi(z)}
\, \biggl(\partial_N \Phi_J \partial^N \Phi^J - \mu^2_J(z) \, \Phi_J
\Phi^J \biggr) \,, 
\en 
where $\mu^2_J(z) = \mu^2_J + g_J \phi(z)$ is
the "dressed" mass due to the interaction of the dilaton with
$\Phi_J$. Note that a similar modification of the mass term of the
string mode dual to the spinor field describing nucleons has been
done in the context of the soft-wall model in
Ref.~\cite{Abidin:2009hr}. The coupling constant $g_J$ will be fixed
later in order to get a massless pion. The metric is defined as 
\eq
ds^2 = \Big(\frac{R}{z}\Big)^2 (\eta_{\mu\nu} dx^\mu dx^\nu  -
dz^2)\,, \hspace*{1cm} 
\eta_{\mu\nu} = {\rm diag}(1, -1, \ldots, -1)
\,. 
\en 
where $R$ is the AdS radius, $g = |{\rm det} g_{MN}| = (R/z)^{2(d+1)}$, 
and $g_{MN}$ is the metric tensor of $d+1$ space.

Next we restrict to the axial gauge $\Phi_{z \ldots}(x,z) = 0$. We
consider the string modes dual to hadrons with total angular momentum $J$,
four--momentum $P$ and propagating in AdS space along
the Poincar\'e coordinates:
\eq
\Phi_{\nu_1 \cdots \nu_J}(x,z) =
\sum\limits_n \ \varphi_{nJ}(z) \ \int \frac{d^d P}{(2\pi)^d}
\ e^{-iPx} \ \epsilon_{\nu_1 \cdots \nu_J}^n(P)
\en
where $\nu_1 \cdots \nu_J$ are the Poincar\'e indices, $n$ is
the radial quantum number and $\epsilon_{\nu_1 \cdots \nu_J}^n(P)$ 
is the polarization tensor. 

Performing the substitution 
\eq 
\varphi_{nJ}(z) =
{\displaystyle{e}}^{\displaystyle{\frac{\phi(z)}{2}}} \, \Big(
\frac{R}{z}\Big)^{J - \frac{d-1}{2}} \, \Phi_{nJ}(z) 
\en 
one can derive the Schr\"odinger--type equation of motion (EOM) for
$\Phi_{nJ}(z)$: 
\eq\label{Eq1} 
\Big[ - \frac{d^2}{dz^2} + U_J(z)
\Big] \Phi_{nJ}(z) = M^2_{nJ} \Phi_{nJ}(z) 
\en 
where $U_J(z)$ is the
effective potential given by 
\eq\label{U_pot} U_J(z) = \kappa^4 z^2
+ \frac{4 a_J^2 - 1}{4z^2} + 2 \kappa^2 \Big( b_J - 1 \Big)
\en
with 
\eq a_J = \frac{1}{2} \, \sqrt{(d - 2J)^2 + 4 (\mu_J R)^2}\,,
\quad\quad b_J = \frac{1}{2} \biggl( g_J R^2 + d - 2J \biggr) \,.
\en 
Analytical solutions of Eq.~(\ref{Eq1}) -- eigenfunctions and eigenvalues
are: 
\eq 
\Phi_{nJ}(z) = \sqrt{\frac{2n!}{(n+a_J)!}} \
\kappa^{1+a_J} \ z^{1/2+a_J} \ e^{-\kappa^2 z^2/2}
L_n^{a_J}(\kappa^2z^2) 
\en 
and 
\eq M^2_{nJ} = 4 \kappa^2 \Big( n + \frac{a_J + b_J}{2} \Big)  \,, 
\en 
where 
\eq 
L_n^m(x) = \frac{x^{-m} e^x}{n!} 
\, \frac{d^n}{dx^n} \Big( e^{-x} x^{m+n} \Big)
\en 
are the generalized Laguerre polynomial.

Restricting to $d=4$ with $(\mu_J R)^2 = 
L^2 - (2 - J)^2$~\cite{Brodsky:2003px}-\cite{Brodsky_ANL} we 
fix the value $g_J R^2 = 4 (J - 1)$ in order to get a massless pion. 
Therefore, in the case $d=4$ we get $a_J = L$ and $b_J = J$, and
the solutions of the Schr\"odinger-type equation read as:  
\eq 
\Phi_{nJ}(z) = \sqrt{\frac{2n!}{(n+L)!}} \ \kappa^{1+L}
\ z^{1/2+L} \ e^{-\kappa^2 z^2/2} L_n^{L}(\kappa^2z^2) 
\en 
and 
\eq  
M^2_{nJ} = 4 \kappa^2 \Big( n + \frac{L + J}{2} \Big) \,. 
\en 
Here we do not divide the total angular momentum $J$ into the quantum numbers 
of the quark--antiquark pair -- orbital angular momentum $L$ and internal 
spin $S$. Such a model ansatz was done in 
Refs.~\cite{Brodsky:2003px}-\cite{Brodsky_ANL}. Because of $J = L$
or $J = L \pm 1$ the present soft-wall model generates linear Regge 
trajectories in both quantum numbers $n$ and $J$ (or $L$): 
$M^2_{nJ} \sim n + J$. Note that the string modes dual to the pseudoscalar
$J^{PC} = 0^{-+}$ and scalar $J^{PC} = 0^{++}$ mesons, and
correspondingly the vector $J^{PC} = 1^{--}$ and axial $J^{PC} =
1^{++}$ mesons, are different from each other (mass spectrum and
wave functions) via the mass parameter of the string mode 
$(\mu_J R)^2$, depending explicitly on the orbital momentum~$L$. 
Inclusion of chiral symmetry breaking effects in the AdS action and their
impact on the hadron properties will be analyzed in the future.

\subsection{AdS and light--front QCD correspondence}

The string mode $\Phi_{nJ}(z)$ can be directly mapped to the LFWF
due to the correspondence of AdS and light-front amplitudes. In
particular, considering the case of two partons $q_1$ and $\bar q_2$
the holographic coordinate $z$ is related to the impact variable
$\zeta$ in the LF formalism as 
\eq 
z \to \zeta, \hspace*{.5cm}\zeta^2 = \bfb^2 x (1 - x)\, 
\en 
where $\bfb$ is the transverse coordinate and
Fourier conjugate to the transverse momentum $\bfk$. In the
massless case we obtain a
relation between the AdS modes and the meson LFWF
$\widetilde{\psi}_{q_1\bar q_2}(x,\zeta)$~\cite{Vega:2009zb}: 
\eq
 \label{Matching_AdS_LF}
 | \widetilde{\psi}_{q_1\bar q_2}(x,\zeta) |^{2} = P_{q_1\bar q_2} \,
 x (1-x) f^2(x) \frac{|\Phi_{nJ}(\zeta)|^{2}}{2\pi\zeta} \,,
\en
where $P_{q_1\bar q_2}$ is the the probability of finding the
valence Fock state $|q_1 \bar q_2\ra$ in the meson $M$:
\eq\label{Probability_cond} 
P_{q_1\bar q_2} = \int\limits_0^1 dx 
\int d^2 \bfb |\widetilde{\psi}_{q_1\bar q_2}(x,\bfb)|^2 \leq 1 \,.
\en 
In the following we restrict to the case of $P_{q_1\bar q_2} =
1$ and only for the pion  we consider $P_{q_1\bar q_2} < 1$ 
(see discussion in Ref.~\cite{Vega:2009zb}). Here $f(x)$ is the
longitudinal mode which is normalized as $\int_0^1 dx f^2(x) = 1$.
In our case $f(x)\equiv 1$. Then the expressions for the meson LFWFs
read: 
\eq\label{psi1} 
\widetilde{\psi}_{q_1\bar q_2}(x,\bfb) =
\sqrt{\frac{2n!}{(n+L)!}} \, \frac{\kappa^{1+L}}{\sqrt{\pi}} \,
|\bfb|^L \, [x(1-x)]^{\frac{1+L}{2}} \, e^{- \frac{1}{2} \kappa^2
x(1-x) \bfb^2} \, L_n^L(\kappa^2 \bfb^2 x(1-x)) \,. 
\en 
The meson LFWF (\ref{psi1}) does not consider massive quarks.
The inclusion of finite quark masses has been considered by us previously
in~\cite{Vega:2009zb}. In particular, the quark masses in the meson
LFWF have been included following a prescription suggested by
Brodsky and de T\'eramond~\cite{Brodsky:2008pg}. Here we illustrate
this procedure for the ground state LFWFs. First one should take the
Fourier transform of~(\ref{psi1}) 
\eq 
\psi_{q_1\bar q_2}(x,\bfk) =
\frac{4\pi}{\kappa \sqrt{x(1-x)}} 
e^{\displaystyle{-\frac{\bfk^2}{2
\kappa^2 x(1-x)}}} \,. 
\en 
In a second step the quark masses are
introduced by extending the kinetic energy of massless quarks with
$K_0 = \frac{\bfk^2}{x(1-x)}$ to the case of massive quarks:
\eq\label{K_mass} 
K_0 \to K = K_0 + m_{12}^2\,, \ \ \ m_{12}^2 =
\frac{m^2_1}{x} + \frac{m^2_2}{1-x} \,. 
\en 
Note that the change proposed in~(\ref{K_mass}) 
is equivalent to the following change of the
kinetic term in the Schr\"odinger EOM: 
\eq 
- \frac{d^{2}}{d
\zeta^{2}} \rightarrow - \frac{d^{2}}{d \zeta^{2}} + m_{12}^2 \,.
\en Finally we obtain \eq\label{psi1_mass}
 \psi_{q_1\bar q_2}(x,\bfk)
= \frac{4\pi N}{\kappa \sqrt{x(1-x)}}
  e^{\displaystyle{-\frac{\bfk^2}{2 \kappa^2 x(1-x)}
- \frac{m_{12}^2}{2\kappa^2}}} \,. 
\en 
As was suggested in~\cite{Teramond_Denver}, 
the quark mass term in the exponential of
Eq.~(\ref{psi1_mass}) can be absorbed in the longitudinal mode for
massive quarks 
\eq 
f(x,m_1,m_2) \equiv N \, f(x) \,
e^{\displaystyle{-\frac{m_{12}^2}{2\kappa^2}}} \,, 
\en 
where $N$ is the normalization constant fixed from 
\eq 
1 = \int\limits_0^1 dx \, f^2(x,m_1,m_2) \,. 
\en 
Hence, the meson LFWFs with massive quarks
can be written down as a product of transverse
$\Phi(\zeta)$, longitudinal $f(x,m_1,m_2)$ and angular $e^{im\phi}$
modes~\cite{Teramond_Denver}: 
\eq 
\widetilde{\psi}_{q_1\bar
q_2}(x,\zeta,m_1,m_2) = \frac{\Phi_{nJ}(\zeta)}{\sqrt{2\pi\zeta}}
f(x,m_1,m_2) \, e^{im\phi}\sqrt{x(1-x)} \,, 
\en 
where $m = 0, \pm 1, \pm 2, \cdots, \pm L$ is the magnetic quantum number. 
One should stress that the way in which massive quarks are introduced 
is not unique. 
In particular, the dimensional parameter entering in the longitudinal mode 
$f(x,m_1,m_2)$ should not necessarily be identified with the parameter  
$\kappa$ characterizing the dilaton field. Later, in the analysis of the  
mass spectrum and the decay constants of heavy--light mesons, we will show  
that the dilaton parameter $\kappa$ should scale as ${\cal O}(1)$ in  
the $1/m_Q$ expansion, where $m_Q$ is the heavy quark mass, while
the dimensional parameter in the longitudinal mode should scale as
${\cal O}(m_Q^{1/2})$. In the case of heavy quarkonia the dimensional
parameter in the longitudinal mode should scale as ${\cal O}(m_Q)$.
Hence, for the longitudinal mode we will use the functional form 
\eq
f(x,m_1,m_2) \equiv N \, f(x) \,
e^{\displaystyle{-\frac{m_{12}^2}{2\lambda_{12}^2}}} \,, 
\en
containing quark masses and an additional scale parameter $\lambda_{12}$.

The meson mass spectrum in the case of massive quarks is given
by~\cite{Teramond_Denver}: 
\eq\label{M2_new} M^2_{nJ} =
\int\limits_0^\infty d\zeta \, \Phi_{nJ}(\zeta) \biggl( - 
\frac{d^2}{d\zeta^2} - \frac{1-4L^2}{4\zeta^2} + \kappa^4 \zeta^2 +
2 \kappa^2 ( J - 1) \biggr) \Phi_{nJ}(\zeta) + \int\limits_0^1 dx
\biggl( \frac{m_1^2}{x} + \frac{m_2^2}{1-x} \biggr) f^2(x,m_1,m_2)
\,. 
\en 
This means that for massive quarks the hadron masses are
shifted due to the last term in the r.h.s. of Eq.~(\ref{M2_new}).
One should stress that the potential in Eq.~(\ref{M2_new}) is not
complete. It includes confinement forces but does not include in
its full context effects of chiral symmetry breaking, which are
important for consistency with the infrared structure of QCD (see
e.g. the discussion in
Refs.~\cite{Karch:2006pv,Colangelo:2007pt,Babington:2003vm,Katz:2005ir,%
Kwee:2007nq,Gherghetta:2009ac,Brodsky:2009zd,Brodsky_ANL}).
Moreover, it does not contain the one--gluon exchange term, which is
sufficient for the description of bottomia hadrons, and also  
hyperfine--splitting terms.
As we stressed before, we intend to include chiral symmetry breaking
in the formalism in a forthcoming work.

\subsection{One--gluon exchange and hyperfine--splitting contributions
to the effective meson potential}

We will include the one--gluon exchange and hyperfine--splitting
terms phenomenologically by
extending the effective potential $U \to U + U_{\rm C} + U_{\rm
HF}$, where $U_{\rm C}$ and $ U_{\rm HF}$ are the contributions of
the color Coulomb and hyperfine (HF) splitting potentials.

Note that, as it was stressed
in~\cite{Sergeenko:1993sn,Gershtein:2006ng}, the trajectories of
bottomia states deviate from linearity. The reason is that, due to
the one--gluon exchange term, there is an additional Coulomb--like
interaction between quarks $V(r) = - 4\alpha_s/3r$, where $\alpha_s$
is the strong coupling constant. Its contribution to the mass
spectrum $M^2$ is negative and proportional to the quark mass
squared~\cite{Gershtein:2006ng,Sergeenko:1993sn}. Therefore, for
light mesons and charmonia states this term can be neglected, while
this is not the case for the bottomia states. Extending the result of
Refs.~\cite{Gershtein:2006ng,Sergeenko:1993sn} to the general case of a
meson containing constituent quarks with masses $m_1$ and $m_2$, we
get the following expression for the shift of $M^2$ due to the color
Coulomb potential: 
\eq 
\Delta M^2_{\rm C} = - \frac{64\alpha_s^2
m_1m_2}{9 \, (n+L+1)^2} \,, 
\en 
where $\alpha_s$ is the QCD coupling, considered as a free parameter. 
The Coulomb potential, which should be included in the effective meson 
potential $U(\zeta)$, reads 
\eq 
U_{\rm C}(\zeta) =  - \frac{\sigma}{\zeta} \,,
\en 
where the coupling constant $\sigma$ is fixed as 
\eq 
\sigma =
\frac{64\alpha_s^2 m_1m_2}{9 \, (n+L+1)^2} \, \biggl\{ \int\limits_0^\infty 
\frac{d\zeta}{\zeta} \, \Phi_{nJ}^2(\zeta) \, \biggr\}^{-1} \,. 
\en 
For the hyperfine--splitting potential $U_{\rm HF}(\zeta)$ one can
use an effective operator containing a free parameter $v$
(softening the original $\delta$-functional form of the
HF-potential) having dimension $M^3$ [see details in
Refs.~\cite{Zhou:2003gj,Karliner:2006fr}]: 
\eq 
U_{\rm HF}(\zeta) = \frac{32\pi\alpha_s}{9} \, \frac{\bm{\sigma_1} \,
\bm{\sigma_2}}{\mu_{12}} \, v \,, 
\en 
where $\bm{\sigma_1}$ and $\bm{\sigma_2}$ are the spin operators of the 
quarks; $\mu_{12} = 2 m_1 m_2/(m_1 + m_2)$. 
Projecting the operator $\bm{\sigma_1} \, 
\bm{\sigma_2}$ between meson states with $S=0,1$ gives 
\eq 
\beta_S =
\la M_S | \bm{\sigma_1} \, \bm{\sigma_2} | M_S \ra = \left\{
\begin{array}{cc}
 -3\,, & S=0 \\
\ 1\,, & S=1 \\
\end{array}
\right . \, .
\en 
Therefore, the mass shift due to the hyperfine--splitting potential is  
\eq 
\Delta M^2_{\rm HF} =
\frac{32\pi\alpha_s}{9} \, \frac{\beta_S \, v}{\mu_{12}} \, .
\en 
Finally, the master formula for meson masses including confinement,
color Coulomb and hyperfine--splitting effects reads: 
\eq M^2_{nJ} =
4 \kappa^2 \Big( n + \frac{L + J}{2} \Big) \, + \, \int\limits_0^1
dx \biggl( \frac{m_1^2}{x} + \frac{m_2^2}{1-x} \biggr)
f^2(x,m_1,m_2) - \frac{64\alpha_s^2 m_1m_2}{9 \, (n+L+1)^2} \, \, +
\,  \frac{32\pi\alpha_s}{9} \, \frac{\beta_S \, v}{\mu_{12}} \,. 
\en 
There are two comments that should be made with respect to further
modifications of the potential $U$. First, as suggested in
Ref.~\cite{Fujita:2009wc} the dilaton scale parameter can be
different for distinct types of mesons -- light and heavy ones. In
particular, we observe that the use of a larger value of $\kappa$ for heavy
mesons helps to improve the description of the mass spectrum and the
leptonic decay constants. Second, in Ref.~\cite{Grigoryan:2010pj} it
was suggested to add a constant term $c^2$ to the effective
potential which is independent on the parameter $\kappa$ and
controls the masses of the ground states. In our formalism such
a constant term in the effective potential can be e.g. generated by
an additional shift of the ``dressed'' mass term 
$\mu_J^2(z) \to \mu_J^2(z) + c^2 z^2/R^2$, 
which leads to the following modification 
of the mass spectrum: $M^2_{nJ} \to M^2_{nJ} + c^2$. Although both of
these modifications can improve the description of meson properties,
their appearance in the AdS action is not well justified. Therefore,
in the present manuscript we do not consider these options and
postpone them for future study.

\section{Properties of light and heavy mesons}

\subsection{Mass spectrum of light mesons}

In the numerical analysis we restrict ourselves to the isospin limit
$m_u = m_d = m$. We fix the free parameters (constituent quark
masses, $\kappa$, $\lambda_{12}$, $\alpha_s$ and $v$) from a fit to
the mass spectrum and the decay constants of light and
heavy mesons. Note that we use a unified value for the dilaton
parameter $\kappa$ for all meson states as dictated by the AdS
action.

The parameters are fixed to the following values. For the 
constituent quark masses we have: 
\eq\label{const_qm} 
m   = 420 \ {\rm MeV}\,, \hspace*{.5cm} 
m_s = 570 \ {\rm MeV}\,, \hspace*{.5cm} 
m_c = 1.6 \ {\rm GeV}\,, \hspace*{.5cm} 
m_b = 4.8  \ {\rm GeV}\,. 
\en
The unified value of the dilaton parameter
is fixed as $\kappa = 550$ MeV for all mesons.
The hyperfine-splitting
parameter has the value $v = 10^{-4}$ GeV$^3$. The strong
coupling $\alpha_s \equiv \alpha_s(\mu_{12}^2)$ depends on the quark
flavor and is consistently calculated using the parametrization of
$\alpha_s$ with ``freezing''~\cite{Badalian:2004xv}:
\eq 
\alpha_s(\mu^2) = \frac{12\pi}{(33 - 2N_f) 
\, {\rm ln}\displaystyle{\frac{\mu^2 + M_B^2}{\Lambda^2}}} 
\en 
where $N_f$ is the number of flavors, 
$\Lambda$ is the QCD scale parameter and $M_B$ is the background mass. 
Choosing $\Lambda = 420$~MeV, $M_B = 854$~MeV 
and using the fixed constituent quark masses from Eq.~(\ref{const_qm})
we obtain the following set of parameters $\alpha_s$: 
\eq 
& &\alpha_s(\mu_{qq}^2) = 0.79\,,  \hspace*{.2cm}
   \alpha_s(\mu_{qs}^2) = 0.77\,,  \hspace*{.2cm}
   \alpha_s(\mu_{ss}^2) = 0.78\,, \hspace*{.2cm}
   \alpha_s(\mu_{qc}^2) = 0.68\,,  \hspace*{.2cm}
   \alpha_s(\mu_{sc}^2) = 0.67\,, \nonumber\\
& &\alpha_s(\mu_{ub}^2) = 0.64\,,  \hspace*{.2cm}
   \alpha_s(\mu_{sb}^2) = 0.61\,, \hspace*{.2cm}
   \alpha_s(\mu_{cc}^2) = 0.52\,, \hspace*{.2cm}
   \alpha_s(\mu_{cb}^2) = 0.42\,, \hspace*{.2cm}
   \alpha_s(\mu_{bb}^2) = 0.33\,,
\en
where $q = u,d$.
The dimensional parameters $\lambda_{12}$ in the longitudinal wave functions
are fitted as:
\eq
& &\lambda_{qq} = 0.63 \ {\rm GeV}\,, \hspace*{.2cm} 
   \lambda_{us} = 1.2  \ {\rm GeV}\,, \hspace*{.2cm}
   \lambda_{ss} = 1.68 \ {\rm GeV}\,, \hspace*{.2cm}
   \lambda_{qc} = 2.5  \ {\rm GeV}\,, \hspace*{.2cm}
   \lambda_{sc} = 3.0  \ {\rm GeV}\,, \nonumber\\
& &\lambda_{qb} = 3.89  \ {\rm GeV}\,, \hspace*{.2cm}
   \lambda_{sb} = 4.18 \ {\rm GeV}\,, \hspace*{.2cm}
   \lambda_{cc} = 4.04 \ {\rm GeV}\,, \hspace*{.2cm}
   \lambda_{cb} = 4.82 \ {\rm GeV}\,, \hspace*{.2cm}
   \lambda_{bb} = 6.77 \ {\rm GeV}\,.
\en 
Here we also already indicate the values used for the heavy-light 
and heavy mesons.
For the probabilities of the ground state pion and kaon we use the 
following 
values: $P_\pi = 0.6$ and $P_K = 0.8$, while for other mesons the
probabilities are supposed to be equal to 1.

The predictions of our approach for the light meson spectrum
according to the $n^{2S+1} L_J$ classification are given in Table~I. 
For the scalar mesons $f_0$ we present results for two limiting cases: 
with nonstrange flavor content 
$f_0[\bar nn] = (\bar u u + \bar d d)/\sqrt{2}$ and with a
strange one $f_0[\bar ss] =  \bar ss $. 

\subsection{Mass spectrum of heavy-light mesons}

Before we apply our approach to the properties of heavy mesons we
would like to check  if our LFWFs are consistent with
model--independent constraints imposed e.g. by heavy quark symmetry,
when the heavy quark mass goes to infinity $m_Q \to \infty$.

The mass spectrum of heavy-light mesons is given by the formula
\eq\label{M_qQ}
M_{qQ}^2 = 4 \kappa^2 \Big(n +\frac{L + J}{2} \Big)
+ \int\limits_0^1 dx \biggl( \frac{m^2_q}{x} + \frac{m_Q^2}{1-x}
\biggr) f^2(x,m_q,m_Q) - \frac{64\alpha_s^2 m_qm_Q}{9 \, (n+L+1)^2}
\, \, + \, \frac{32\pi\alpha_s}{9} \, \frac{\beta_S \, v}{\mu_{qQ}}
\,.
\en
The longitudinal mode for heavy--light mesons is of the form
\eq
f(x,m_q,m_Q) \equiv N \, f(x) \,
e^{\displaystyle{-\frac{m_{qQ}^2}{2\lambda_{qQ}^2}}} \,, 
\en 
where
$\lambda_{qQ}$ is the dimensional parameter which scales as 
${\cal O}(m_Q^{1/2})$. In the following, for convenience, we express
$\lambda_{qQ}$ as $\lambda_{qQ}^2 = m_qm_Q/r$, where $r$ is a
parameter of order ${\cal O}(1)$. The scaling of the parameter
$\kappa \sim {\cal O}(1)$ is fixed by the scaling law of the
leptonic constants of heavy--light mesons in the heavy quark limit
(see discussion in Sec.~\ref{dec_const}). This behavior of
$\kappa$ is also consistent with the mass spectroscopy of
heavy--light mesons constrained by heavy quark effective
theory~(HQET)~\cite{Neubert:1993mb}. In particular, the $1/m_Q$
expansion of their masses should be
\eq 
M_{qQ} = m_Q + \bar\Lambda + {\cal O}(1/m_Q) \,,
\en
where the scale parameter $\bar\Lambda$ is
of order ${\cal O}(1)$, and the mass splitting of vector and
pseudoscalar states $\Delta M_{qQ} = M_{qQ}^V - M_{qQ}^P$
should be of order $1/m_Q$: 
\eq 
\Delta M_{qQ} = \frac{2}{M_{qQ}^V +
M_{qQ}^P} \, \biggl( \kappa^2 + \frac{64 \pi \alpha_s}{9} \,
\frac{\beta_S \, v}{m_q}\biggr) \sim \frac{1}{m_Q} \,. 
\en 
The mass
splitting $\Delta M_{qQ}$ gets contributions from two sources ---
confinement and the hyperfine--splitting potential. Both
contributions are of order ${\cal O}(1)$ in the heavy quark mass
expansion. In Appendix~\ref{HQL_int} we evaluate the r.h.s. of
Eq.~(\ref{M_qQ}) and give an expression for the scale parameter
$\bar\Lambda$.

Numerical values for the charm and bottom heavy-light mesons with
different spin-parity are given in Table~II according to the
$n^{2S+1} L_J$ classification. 
The four columns with the results
correspond to the variation of $L=0,1,2,3$. Data for
the ground states are given in the brackets.

\subsection{Mass spectrum of heavy quarkonia}

For the mass spectrum of heavy quarkonia $(Q_1 \bar Q_2)$ we present
our results in Tables~III. Again it is interesting to consider
the limit of heavy quark masses. Here we follow the idea suggested
in~\cite{Teramond_Denver}, and express the longitudinal momentum
fractions through the $z$-component of the internal momentum
${\bf k} = (\bfk, k_z)$ as (see also~\cite{Jaus:1989au}):
\eq
x = \frac{e_1 + k_z}{e_1 + e_2}\,, \hspace*{.25cm} 1 - x = \frac{e_2 -
k_z}{e_1 + e_2}\,, \hspace*{.25cm} 
\en 
where $e_i = \sqrt{m_{Q_i}^2
+ {\bf k}^2}$ and ${\bf k}^2 = \bfk^2 + k_z^2$. When considering the
heavy quark limit $m_{Q_1}, m_{Q_2} \gg \bfk, k_z$ we get 
\eq 
x = \frac{m_{Q_1} + k_z}{m_{Q_1} + m_{Q_2}} + {\cal O}(1/m_Q^2)\,,
\hspace*{.25cm} 1 - x = \frac{m_{Q_2} - k_z}{m_{Q_1} + m_{Q_2}} +
{\cal O}(1/m_Q^2) \,. 
\en 
Hence, we have 
\eq \frac{m_{Q_1}^2}{x} +
\frac{m_{Q_2}^2}{1 - x} = (m_{Q_1} + m_{Q_2})^2 + {\cal O}(1) \,.
\en 
Therefore, the leading term of the integral containing the
longitudinal mode is simply given by $(m_{Q_1} + m_{Q_2})^2$ which
is the leading contribution to the mass squared of the heavy
quarkonia. This means that we correctly reproduce an expansion of the
heavy quarkonia mass in the heavy quark limit: 
\eq 
M_{Q_1\bar Q_2} =
m_{Q_1} + m_{Q_2} + E + {\cal O}(1/m_{Q_{1,2}}) \,, 
\en 
where $E$ is the binding energy. Numerical values for the quarkonia masses 
are shown in Table III, according to the $n^{2S+1} L_J$ classification. 

A graphical summary of our results is given in Figs.1-6 where we display 
the calculated meson mass spectra for light (Figs.1 and 2), heavy-light 
(Figs.3 and 4) and heavy quarkonia (Figs.5 and 6). Because of the inclusion 
of the color Coulomb potential the lines are bent down for low values of 
$L$ or $n$ and therefore deviate from the linear behavior of 
Regge-trajectories. We indicate the available data (central values 
by black circles and sizable error bars by vertical lines). 
Here we find that the calculated mass spectra 
are in agreement with predictions of other holographic models 
(see e.g. Refs.~\cite{Erlich:2005qh,Da Rold:2005zs,Karch:2006pv}). 
The improvement of the present approach is that a consistent description 
of light, heavy-light and double-heavy mesons is achieved within 
the same holographic model. For comparison, in the literature so far 
different types of mesons have been considered separately. 
As already mentioned, we use a universal value for the dilaton parameter 
$\kappa$. However, adapted values of $\kappa$ for different types of mesons 
leads of course to a better fit to the data.

\subsection{Leptonic and radiative meson decay constants}
\label{dec_const}

In the following we define further fundamental quantities of
mesons, which are calculated in the present paper --
decay constants of pseudoscalar ($f_P$) 
\eq\label{matrix_el_fP} 
\la 0 | \bar q_2(0) \gamma^\mu \gamma^5 q_1(0) | M_P(P) \ra = i P^\mu \,
f_P \,, \en and vector ($f_V$) mesons \eq\label{matrix_el_fV} \la 0
| \bar q_2(0) \gamma^\mu q_1(0) | M_V(P,\lambda) \ra = \epsilon^\mu
(P,\lambda) \, M_V \, f_V \,. 
\en 
The definitions of the meson Fock states are given 
in Appendix~\ref{app:LF_Fock_states}.

For convenience we determine the decay constants restricting
to $\mu = +$ and spin projection $\lambda=0$ for the
vector mesons. In this case Eqs.~(\ref{matrix_el_fP}) and
(\ref{matrix_el_fV}) are simply reduced to 
\eq 
& &\big\la0\big| \bar q_2(0) \gamma^+ \gamma^5 q_1(0) \big|P\big\ra
=P^+ f_P\,,\\ 
& &\big\la0\big| \bar q_2(0) \gamma^+ q_1(0)
\big|V(P,\lambda=0)\big\ra =P^+f_V\,. 
\en 
Finally, we get for the couplings of the ground state mesons: 
\eq\label{fP_result} 
f_P &=&
f_V \ = \ 2\sqrt{6}\int\limits_0^1dx\int\frac{d^2\bfk}{16\pi^3}
\psi_{q_1 \bar q_2}(x,\bfk) \, f(x,m_1,m_2) = \sqrt{\frac{6}{\pi}}
\, \int\limits_0^1dx \,
\tilde\psi_{q_1\bar q_2}(x,\bfb=0) \, f(x,m_1,m_2) \nonumber\\
&=& \kappa \frac{\sqrt{6}}{\pi} \int\limits_0^1dx \sqrt{x(1-x)} \,
 f(x,m_1,m_2)
\label{eq:fp_fv} 
\en 
In the case of massless quarks $f_P$ and $f_V$
are proportional to the dilaton scale parameter $\kappa$: 
\eq 
f_P = f_V =\frac{\kappa \sqrt6}{8}\,. 
\en 
In the heavy quark limit ($m_Q \to \infty$) the scaling of the
leptonic decay constants of heavy--light mesons is in
agreement with HQET: 
\eq 
f_P^{\rm HQL} = f_V^{\rm HQL} = \kappa  \frac{\sqrt{6}}{\pi}
\sqrt{\frac{m_q}{m_Q}} \, \frac{\int\limits_0^\infty dz \,
e^{-\frac{r}{2}(z + \frac{1}{z})}} {\bigg[\int\limits_0^\infty dz \,
e^{-r(z + \frac{1}{z})}\bigg]^{1/2}} \ \sim \frac{1}{\sqrt{m_Q}} \,.
\en 
Again, as in the case of the mass spectrum of heavy--light mesons,
it is sufficient to propose the following scaling of dimensional
parameters in our holographic approach: $\kappa \sim {\cal O}(1)$
and $\lambda_{qQ} \sim {\cal O}(\sqrt{m_Q})$.

When dealing with vector mesons with hidden flavor one should also include
the flavor factor $c_V$
\eq
c_V=\left\{
\begin{array}{ll}
1/\sqrt2\,,&V=\rho^0\\
2/3\,,&V=J/\psi\\
1/3\,,&V=\phi,\Upsilon\\
1/(3\sqrt2)\,,&V=\omega
\end{array}
\right. \en which arise from the flavor structure of the vector mesons
\begin{align}
\rho^0&=\frac{1}{\sqrt2}\big(\bar u u - \bar d d\big),
     &\omega&=\frac{1}{\sqrt2}\big(\bar u u + \bar d d\big),
&\phi&=-\bar s s,& J/\psi&=\bar c c, &\Upsilon&=-\bar b b
\end{align}
and the structure of the corresponding electromagnetic quark currents
\eq
V_{\rho,\omega}^\mu&=&e_u\bar u\gamma^\mu u+e_d\bar d\gamma^\mu d\\
V_{\phi,J/\psi,\Upsilon}^\mu&=&e_q\bar q\gamma^\mu q,\text{ with
$q=s,c,b$}\,. 
\en 
Our results for $f_P$ and $f_V$ are presented in
Tables~IV-VI. Note that with the universal value of the dilaton
scale parameter $\kappa = 550$ MeV the data for
the coupling constants of the light mesons can be well reproduced.
For heavy--light mesons we
need a bit larger value for the parameter $\kappa$, because the
leptonic decay constants are proportional to $\kappa$. For the
description of the leptonic decay constants of heavy quarkonia states we need
an even larger value for $\kappa$. In particular, it should be
roughly  2, 3 and 4 times larger for $c \bar c$, $c \bar b$ and $b
\bar b$ states, respectively, than the unified value of 550 MeV.

\section{Conclusions}

In conclusion, we present an analysis of the mass
spectrum and decay properties of light, heavy--light mesons and
heavy quarkonia in an holographic soft-wall model using conventional 
sign of the dilaton profile with $\phi(z) = \kappa^2 z^2$. 
In our calculations we 
consider in addition one-gluon exchange and hyperfine--splitting  
corrections phenomenologically 
by modifying the potential. We show 
that the obtained results for heavy--light mesons are consistent with 
constraints imposed by HQET. In future work we plan to improve 
the description of meson data and also to extend our formalism to baryons. 

\begin{acknowledgments}

The authors thank Stan Brodsky, Guy de T\'eramond, Herry Kwee and
Oleg Andreev for useful discussions. This work was supported by the DFG 
under Contract No. FA67/31-2 and No. GRK683, 
by FONDECYT (Chile) under Grant No. 1100287. This research is also part 
of the European Community-Research Infrastructure Integrating Activity 
``Study of Strongly Interacting Matter'' (HadronPhysics2, Grant Agreement 
No. 227431), Russian President grant ``Scientific Schools'' No. 3400.2010.2, 
Federal Targeted Program "Scientific and scientific-pedagogical personnel 
of innovative Russia" Contract No. 02.740.11.0238. 
A. V. acknowledges the financial support from FONDECYT (Chile) 
Grant No. 3100028. 

\end{acknowledgments}

\appendix\section{Light--front meson Fock states}\label{app:LF_Fock_states}

Here we define the Fock states of pseudoscalar $(P)$, scalar
$(S)$, vector $(V)$ and axial $(A)$ mesons, restricting to the
valence quark-antiquark contribution only. The corresponding mesonic
eigenstates with momentum $P = (P^+, P^-, \bfP)$ are given by 
\eq\label{eigenstateMP}
\hspace*{-.4cm} \big|M_{P,S}(P)\big\ra = \frac{2P^+}{\sqrt{2 N_c}}
\int\limits_0^1 dx \int\frac{d^2\bfk}{16\pi^3} \psi_{q_1\bar
q_2}(x,\bfk) \Big[d^{\dagger a}_{q_1\uparrow}(p_1)
  b^{\dagger a}_{q_2\downarrow}(p_2)
- d^{\dagger a}_{q_1\downarrow}(p_1)
  b^{\dagger a}_{q_2\uparrow}(p_2) \Big] |0\ra
\en
and
\eq\label{eigenstateMV}
\big|M_{V,A}(P,\lambda)\big\ra = \frac{2P^+}{\sqrt{2N_c}}
\int\limits_0^1 dx \int\frac{d^2\bfk}{16\pi^3}
\psi_{q_1\bar q_2}(x,\bfk) \,
\left\{
\begin{array}{ll}
d^{\dagger a}_{q\uparrow}(xP)\,b_{q\uparrow}^{\dagger a}((1-x)P)\big|0\big>
\sqrt{2}\,, &\lambda=+1\\
d^{\dagger a}_{q\uparrow}(xP)\,b_{q\downarrow}^{\dagger a}((1-x)P)
+d^{\dagger a}_{q\downarrow}(xP)\,
b_{q\uparrow}^{\dagger a}((1-x)P)\big|0\big>\,,
&\lambda=0\\
d^{\dagger a}_{q\downarrow}(xP)\,
b_{q\downarrow}^{\dagger a}((1-x)P)\big|0\big>
\sqrt{2}\,, &\lambda=-1
\end{array}
\right. 
\en 
with 
\eq (p_1) = (xP^+,x\bfP+\bfk)\,, \hspace*{.5cm}
(p_2) = ((1-x)P^+,(1-x)\bfP-\bfk)\,. \nonumber 
\en 
Here $a$ is the 
color index and $N_c = 3$ is the number of colors. 
In the expansion~(\ref{eigenstateMP}) of the meson state over the basis of
noninteracting n-particles states we restrict to the two-quark
valence state. Notice that above formulas can be extended easily
to the case of mesons with more nontrivial flavor structure, e.g.
including mixing of nonstrange and strange quark components. Also
one can consider pure glueball states and their mixing with
quarkonia.

The operators $b(d)$ obey the anticommutation relations:
\eq\label{anticommut}
\big\{b^{a}(p),b^{\dagger\,a^\prime}(p^\prime\big\}
=\big\{d^{a}(p),d^{\dagger\,a^\prime}(p^\prime\big\}
=(2\pi)^3 \delta^{aa^\prime} \delta(p^+-p^{\prime+})
\delta^{(2)}(\bfp-\bfppr) \,.
\en
The states $|M_H(P)\ra$ and momentum LFWF are normalized according to
\eq
\!\! \la M_P(P^\prime) | M_P(P)\ra = 2 P^+ (2\pi)^3
\delta(P^+ \! - \! P^{\prime+})
\delta^{(2)}(\bfP \! - \! \bfPpr)\,,
\en
\eq
\!\! \la M_V(P^\prime,\lambda^\prime) | M_V(P,\lambda)\ra = 2 P^+ (2\pi)^3
\delta(P^+ \! - \! P^{\prime+})
\delta^{(2)}(\bfP \! - \! \bfPpr) \delta_{\lambda\lambda^\prime}\,,
\en
and
\eq\label{norm_cond}
\int\limits_0^1 dx \int \frac{d^2 \bfk}{16\pi^3}
|\psi_{\bar q_2 q_1}(x,\bfk)|^2 = 1 \, .
\en
The polarization vectors $\epsilon^\mu (P,\lambda)$
in the light-cone representation read as
\eq
\epsilon^\mu(P,\lambda)=\left\{
\begin{array}{ll}
\biggl(\frac{P^+}{M_V},\frac{\bfP^{\,2}-M_V^2}{M_VP^+},
\frac{\bfP}{M_V}\biggr)\,,& \ \ \lambda=0\\[3mm]
\biggl(0,\frac{2\bfe(\pm1)\bfP}{P^+},\bfe(\pm1)\biggr)\,,& \ \ \lambda=\pm1
\end{array}
\right.
\en
with $\bfe(\pm1)=\mp\frac{(1, \pm i)}{\sqrt2}$.

Note that the normalization condition~(\ref{norm_cond}) is
approximately valid only for mesons, where the two-parton
quark-antiquark component is dominant. As we stressed before (also
see discussion in Ref.~\cite{Vega:2009zb}), this is not the case for
the pion, where the probability of the valence quark-antiquark component is
less than 1 [see also Eq.~(\ref{Probability_cond})].

\section{Evaluation of integrals
in the heavy quark limit}\label{HQL_int}

We evaluate the integral in the r.h.s. of Eq.~(\ref{M_qQ}) with
\eq 
J = \int\limits_0^1 dx \biggl( \frac{m^2_q}{x} + \frac{m_Q^2}{1-x}
\biggr) \, f^2(x,m_q,m_Q) = \frac{\int\limits_0^1 dx \biggl(
\frac{m^2_q}{x} + \frac{m_Q^2}{1-x} \biggr)
\exp\biggl(-\frac{m^2_q}{\lambda_{qQ}^2x} +
\frac{m_Q^2}{\lambda_{qQ}^2(1-x)} \biggr)} {\int\limits_0^1 dx
\exp\biggl(-\frac{m^2_q}{\lambda_{qQ}^2x} +
\frac{m_Q^2}{\lambda_{qQ}^2(1-x)} \biggr)} \,.  
\en 
Scaling the variable
$x=z m_q/m_Q$ and then performing an expansion in powers of $1/m_Q$: 
\eq 
J = m_Q^2 + m_Q m_q \frac{\int\limits_0^\infty dz \Big(z +
\frac{1}{z}\Big) \exp\biggl(-\frac{m_qm_Q}{\lambda_{qQ}^2} \Big(z +
\frac{1}{z}\Big) \biggr)} {\int\limits_0^\infty dz
\exp\biggl(-\frac{m_qm_Q}{\lambda_{qQ}^2} \Big(z + \frac{1}{z}\Big)
\biggr)} + {\cal O}(1) 
\en 
where the parameter $\lambda_{qQ}$ scales as $\sqrt{m_Q}$. 
Such a scaling of $\kappa$ is consistent with the
scaling of the leptonic coupling constants of heavy--light mesons (see
Sec.~\ref{dec_const}). For convenience, we introduce the parameter
$r=m_qm_Q/\lambda_{qQ}^2$. Finally, the expansion of the heavy--light
meson mass reads 
\eq 
M_{qQ} = m_Q + \bar\Lambda + {\cal O}(1/m_Q) 
\en
where 
\eq 
\bar\Lambda = m_q I \en and \eq I = \frac{1}{2} \,
\frac{\int\limits_0^\infty dz \Big(z + \frac{1}{z}\Big) \, e^{-r(z +
\frac{1}{z})}} {\int\limits_0^\infty dz \, e^{-r(z + \frac{1}{z})}} \,. 
\en

\newpage

\begin{table}
\caption{Masses of light mesons \label{tab:2}}
\begin{tabular}{|l|c|c|c|l|l|l|l|}
\hline
Meson&$n$&$L$&$S$&\multicolumn{4}{c|}{Mass [MeV]} \\
\hline
$\pi$&0&0,1,2,3&0&$M_{\pi(140)}=140$&$M_{b_1(1235)}=1355$
&$M_{\pi_2(1670)}=1777$&$M_{\pi_4}=2099$ \\ \hline
$\pi$&0,1,2,3&0&0&$M_{\pi(140)}=140$&$M_{\pi(1300)}=1355$
&$M_{\pi(1800)}=1777$&$M_{\pi(4s)}=2099$ \\ \hline
$K$& 0&0,1,2,3&$\;0\;$&$M_{K}=495$&$ M_{K_1(1270)}=1505$
&$ M_{K_2(1770)}=1901$ & $M_{K_3}=2207$ \\ \hline
$\eta$&0,1,2,3&0&0&$M_{\eta(1s)}=544$
&$ M_{\eta(2s)}=1552$&$ M_{\eta(3s)}=1946$
&$ M_{\eta(4s)}=2248$ \\ \hline
$f_0[\bar n n]$&0,1,2,3&1&1&$M_{f_0(1p)}=1114$&$M_{f_0(2p)}=1600$
&$M_{f_0(3p)}=1952$&$M_{f_0(4p)}=2244$ \\ \hline
$f_0[\bar s s]$&0,1,2,3&1&1&$M_{f_0(1p)}=1304$&$M_{f_0(2p)}=1762$
&$M_{f_0(3p)}=2093$&$M_{f_0(4p)}=2372$ \\ \hline
$a_0(980)$&0,1,2,3&1&1&$M_{a_0(1p)}=1114$&$M_{a_0(2p)}=1600$
&$M_{a_0(3p)}=1952$&$M_{a_0(4p)}=2372$ \\ \hline
$\rho(770)$&0,1,2,3&0&1&$M_{\rho(770)}=804$&$M_{\rho(1450)}=1565$
&$M_{\rho(1700)}=1942$&$M_{\rho(4s)}=2240$ \\ \hline
$\rho(770)$&0&0,1,2,3&1&$M_{\rho(770)}=804$&$M_{a_2(1320)}=1565$
&$M_{\rho_3(1690)}=1942$&$M_{a_4(2040)}=2240$ \\ \hline
$\omega(782)$&0,1,2,3&0&1&$M_{\omega(782)}=804$&$M_{\omega(1420)}=1565$
&$M_{\omega(1650)}=1942$&$M_{\omega(4s)}=2240$ \\ \hline
$\omega(782)$&0&0,1,2,3&1&$M_{\omega(782)}=804$&$M_{f_2(1270)}=1565$
&$M_{\omega_3(1670)}=1942$&$M_{f_4(2050)}=2240$ \\ \hline
$\phi(1020)$ &0,1,2,3&0&1&$M_{\phi(1s)}=1019$&$M_{\phi(2s)}=1818$
&$M_{\phi(3s)}=2170$&$M_{\phi(4s)}=2447$ \\ \hline
$a_1(1260)$&0,1,2,3&1&1&$M_{a_1(1p)}=1358$&$M_{a_1(2p)}=1779$
&$M_{a_1(3p)}=2101$&$M_{a_1(4p)}=2375$ \\ \hline
\end{tabular}

\vspace*{.5cm}

\caption{Masses of heavy--light mesons \label{tab:3}}
\begin{tabular}{|l|c|c|c|c|c|c|c|c|}
\hline
Meson&$J^{\rm P}$&$n$&$L$&$S$&\multicolumn{4}{c|}{Mass [MeV]} \\
\hline
$D(1870)$&$0^{-}$&0&0,1,2,3         &0& 1857 & 2435 & 2696 & 2905 \\ \hline
$D^{\ast}(2010)$&$1^{-}$&0&0,1,2,3  &1& 2015 & 2547 & 2797 & 3000 \\ \hline
$D_s(1969)$&$0^{-}$&0&0,1,2,3       &0& 1963 & 2621 & 2883 & 3085 \\ \hline
$D^{\ast}_s(2107)$&$1^{-}$&0&0,1,2,3&1& 2113 & 2725 & 2977 & 3173 \\ \hline
$B(5279)$&$0^{-}$&0&0,1,2,3         &0& 5279 & 5791 & 5964 & 6089 \\ \hline
$B^{\ast}(5325)$&$1^{-}$&0&0,1,2,3  &1& 5336 & 5843 & 6015 & 6139 \\ \hline
$B_s(5366)$&$0^{-}$&0&0,1,2,3       &0& 5360 & 5941 & 6124 & 6250 \\ \hline
$B^{\ast}_s(5413)$&$1^{-}$&0&0,1,2,3&1& 5416 & 5992 & 6173 & 6298 \\ \hline
\end{tabular}

\vspace*{.5cm}

\caption{Masses of heavy quarkonia $c\bar c$, $b\bar b$ and
$c \bar b$}
\label{tab5}
\begin{tabular}{|l|c|c|c|c|l|l|l|l|}
\hline
Meson&$J^{\rm P}$&$n$&$L$&$S$&\multicolumn{4}{c|}{Mass [MeV]} \\ \hline
$\eta_c(2980)$&$0^{-}$ &0,1,2,3&0&0  & 2997 & 3717 & 3962 & 4141 \\\hline
$\psi(3097))$ &$1^{-}$ &0,1,2,3&0&1  & 3097 & 3798 & 4038 & 4213 \\ \hline
$\chi_{c0}(3415)$&$0^{+}$&0,1,2,3&1&1& 3635 & 3885 & 4067 & 4226 \\ \hline
$\chi_{c1}(3510)$&$1^{+}$&0,1,2,3&1&1& 3718 & 3963 & 4141 & 4297 \\ \hline
$\chi_{c2}(3555)$&$2^{+}$&0,1,2,3&1&1& 3798 & 4038 & 4213 & 4367 \\ \hline
$\eta_{b}(9390)$&$0^{-}$ &0,1,2,3&0&0& 9428 & 10190& 10372& 10473 \\ \hline
$\Upsilon(9460)$&$1^{-}$ &0,1,2,3&0&1& 9460 & 10219& 10401& 10502\\ \hline
$\chi_{b0}(9860)$&$0^{+}$&0,1,2,3&1&1&10160 & 10343& 10444& 10521 \\ \hline
$\chi_{b1}(9893)$&$1^{+}$&0,1,2,3&1&1&10190 & 10372& 10473& 10550 \\ \hline
$\chi_{b2}(9912)$&$2^{+}$&0,1,2,3&1&1&10219 & 10401& 10502& 10579 \\ \hline
$B_c(6276)$&$0^{-}$      &0,1,2,3&0&0& 6276 & 6911 & 7092 & 7209  \\ \hline
\end{tabular}
\end{table}

\begin{table}
\def\arraystretch{1.25}
\caption{Decay constants $f_P$ of pseudoscalar mesons in MeV}\label{tab8}
\begin{tabular}{|l|c|c|}
\hline
Meson &Data~\cite{Nakamura:2010zz} & \ Our \ \\ \hline
$\pi^-$& $130.4\pm 0.03 \pm 0.2$ & 131 \\
\hline
$K^-$ & $156.1\pm 0.2 \pm 0.8$ & 155 \\
\hline
$D^+$ & $206.7 \pm 8.9$ & 167 \\
\hline
$D_s^+$ & $257.5 \pm 6.1$ & 170 \\
\hline
$B^-$&$193 \pm 11$ & 139 \\
\hline
$B_s^0$&$253 \pm 8 \pm 7$ & 144 \\
\hline
$B_c$& $489\pm 5 \pm 3$~\cite{Chiu:2007bc} & 159 \\
\hline
\end{tabular}

\vspace*{1cm}

\def\arraystretch{1.25}
\caption{Decay constants $f_V$ of vector mesons with open flavor in MeV}
\label{tab10}
\begin{tabular}{|l|c|c|c|l|}
\hline
Meson &Data & \ Our \ \\ \hline
$\rho^+$ & $210.5 \pm 0.6$~\cite{Nakamura:2010zz} & 170 \\ \hline
$D^\ast$&$245\pm 20^{+3}_{-2}$~\cite{Becirevic:1998ua}& 167 \\ \hline
$D_s^\ast$&$272\pm 16^{+3}_{-20}$~\cite{Aubin:2005ar}& 170 \\ \hline
$B^\ast$&$196\pm24^{+39}_{-2}$~\cite{Becirevic:1998ua}& 139 \\ \hline
$B^\ast_s$&$229\pm20^{+41}_{-16}$~\cite{Becirevic:1998ua}& 144 \\ \hline
\end{tabular}

\vspace*{1cm}

\def\arraystretch{1.25}
\caption{Decay constants $f_V$ of vector mesons with hidden flavor in MeV}
\label{tab12a}
\begin{tabular}{|l|c|c|c|}
\hline
Meson &Data~\cite{Nakamura:2010zz} & \ Our \ \\ \hline 
$\rho^0$& 154.7 $\pm$ 0.7 & 120  \\ \hline
$\omega$ & 45.8 $\pm$ 0.8 & 40   \\ \hline
$\phi$   & 76   $\pm$ 1.2 & 58 \\ \hline
$J/\psi$ &277.6 $\pm$ 4   & 116  \\ \hline
$\Upsilon(1s)$& 238.5 $\pm$ 5.5 & 56 \\ \hline
\end{tabular}
\end{table}

\newpage 

\begin{figure}[hbp]
\includegraphics[scale=0.6,angle=0]{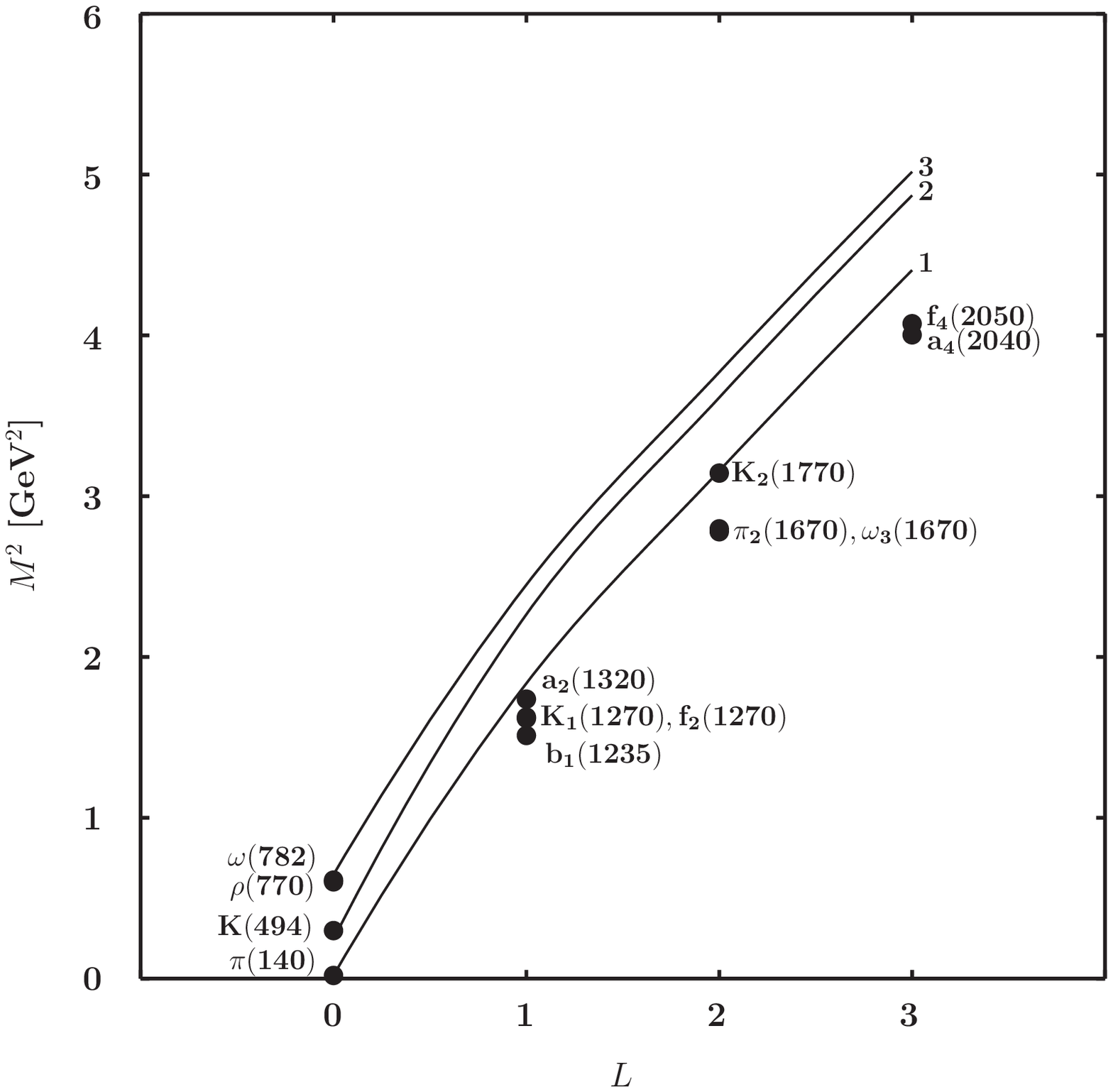}
\caption{Mass spectrum of $\pi$, $K$, $\rho$ and $\omega$ 
mesons in dependence on $L$. 
Central values of data are indicated by black circles. 
The numbers mark the corresponding family trajectories:  
1 - $\pi$ mesons; 
2 - $K$ mesons; 
3 - $\rho, \omega$ mesons. 
} 
\label{fig1}

\vspace*{.75cm}

\includegraphics[scale=0.6,angle=0]{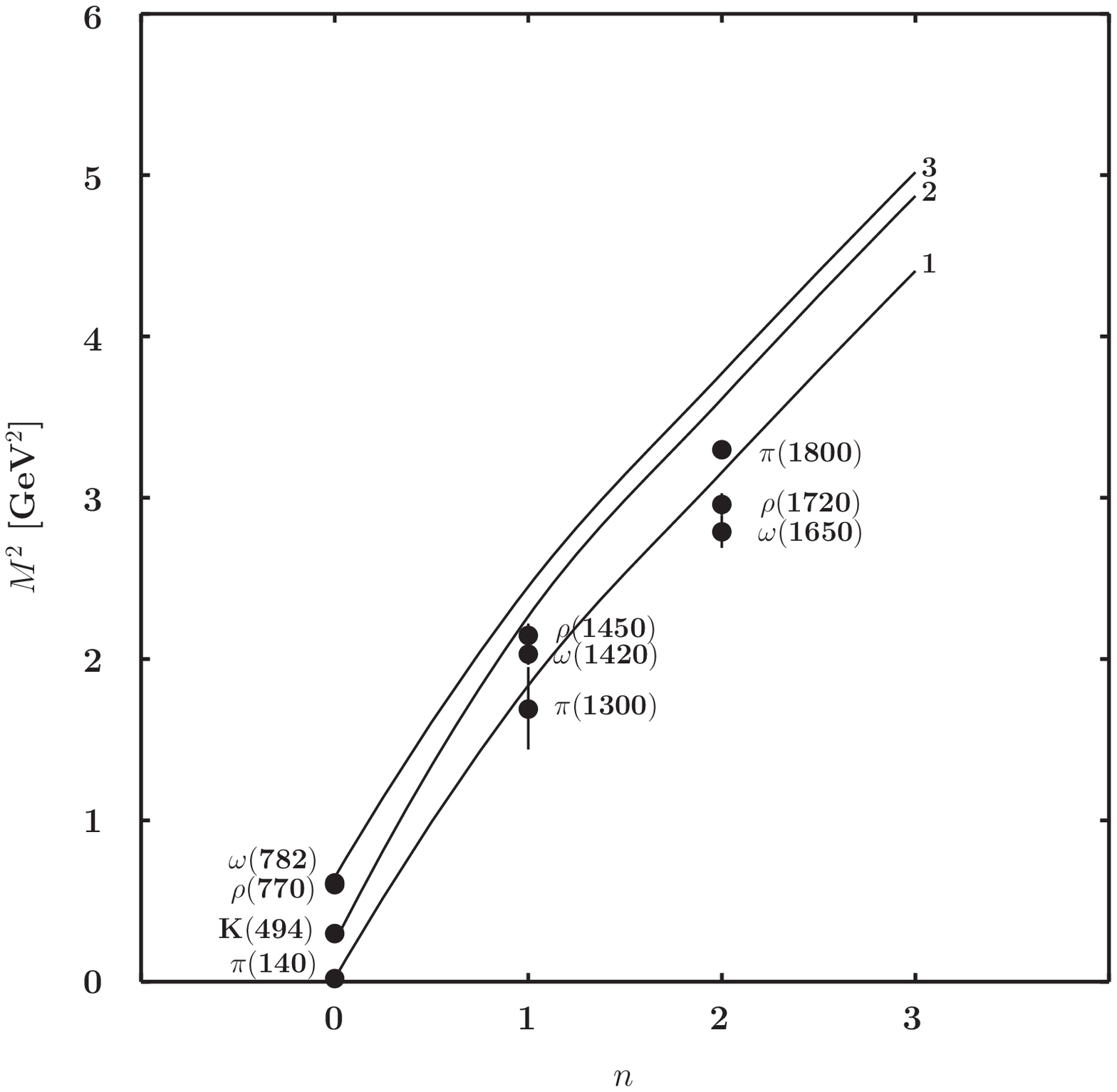}
\caption{Mass spectrum of $\pi$, $K$, $\rho$ and $\omega$ 
mesons in dependence on $n$. 
Central values of data are indicated by black circles 
(for a few states error bars are included when they are sizeable). 
The numbers mark the corresponding family trajectories: 
1 - $\pi$ mesons; 
2 - $K$ mesons; 
3 - $\rho, \omega$ mesons. 
}
\label{fig2}
\end{figure}

\newpage 

\begin{figure}[hbp]
\includegraphics[scale=0.6,angle=0]{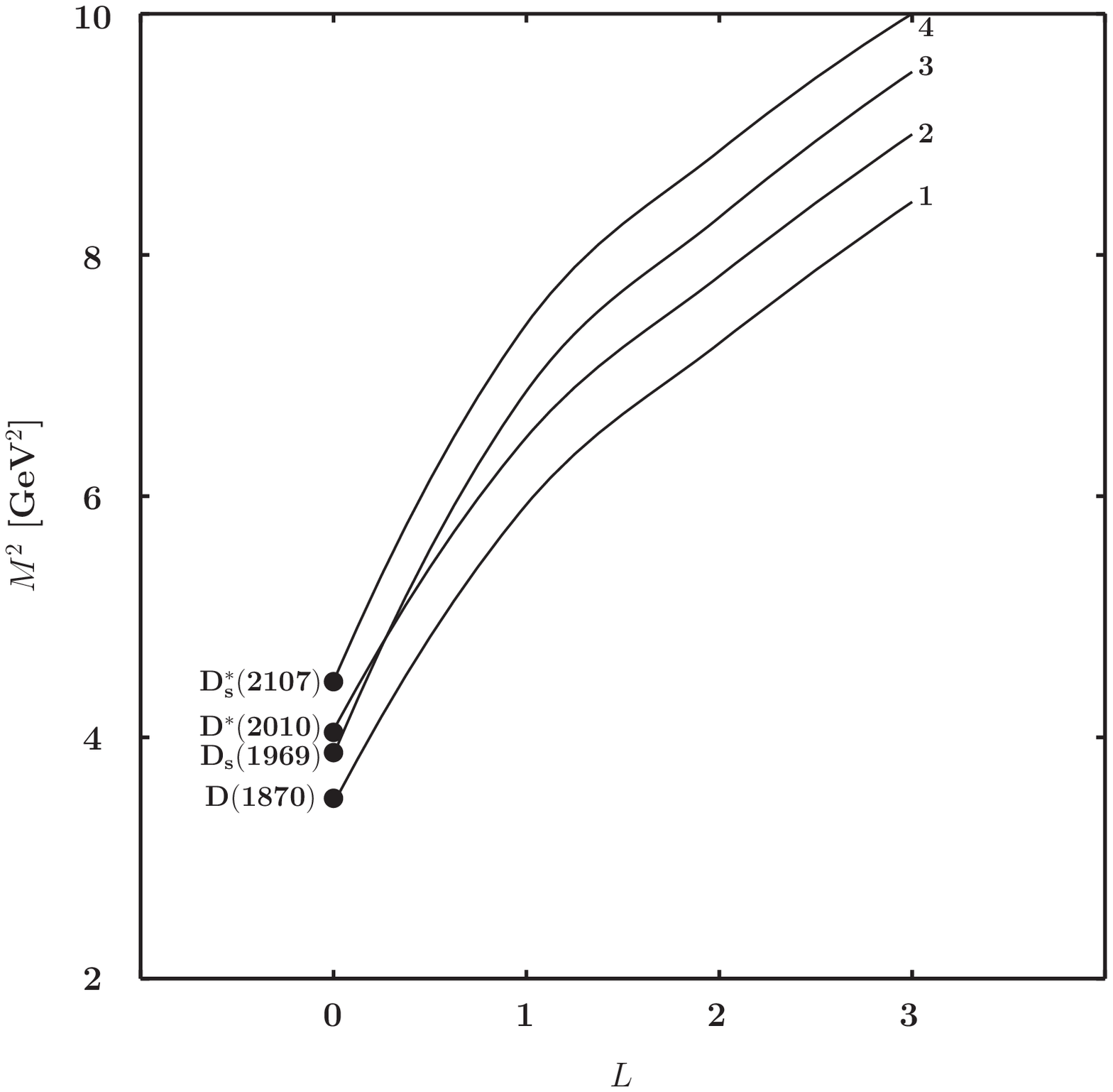}
\caption{Mass spectrum of charmed mesons in dependence on $L$. 
Central values of data are indicated by black circles. 
The numbers mark the corresponding family trajectories: 
1 - $D$ mesons; 
2 - $D^\ast$ mesons; 
3 - $D_s$ mesons;
4 - $D^\ast_s$ mesons. 
} 
\label{fig3}

\vspace*{.75cm}

\includegraphics[scale=0.6,angle=0]{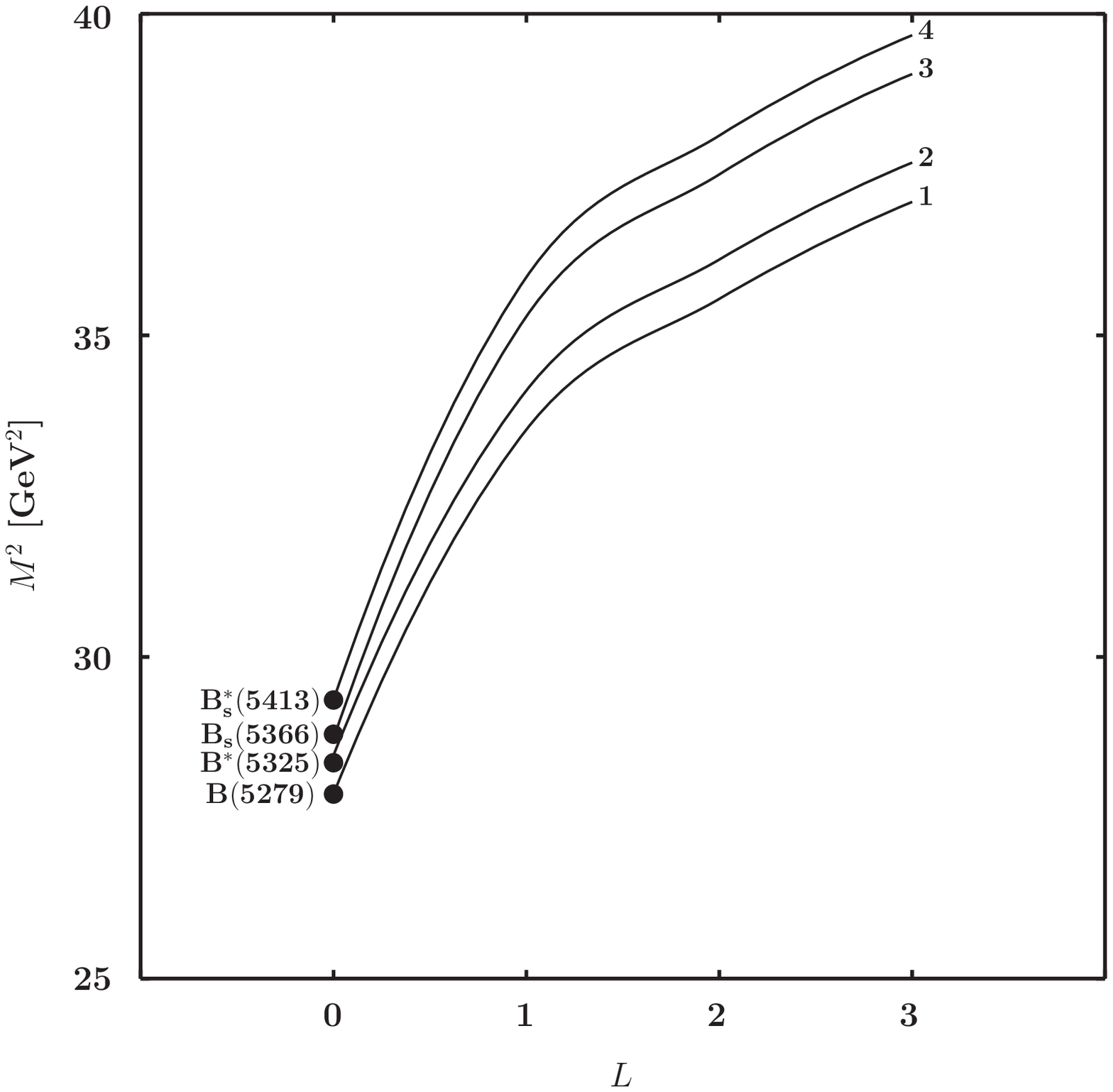}
\caption{Mass spectrum of bottom mesons in dependence on $L$. 
Central values of data are indicated by black circles. 
The numbers mark the corresponding family trajectories: 
1 - $B$ mesons; 
2 - $B^\ast$ mesons; 
3 - $B_s$ mesons;
4 - $B^\ast_s$ mesons. 
}
\label{fig4}
\end{figure} 

\begin{figure}[hbp]
\includegraphics[scale=0.6,angle=0]{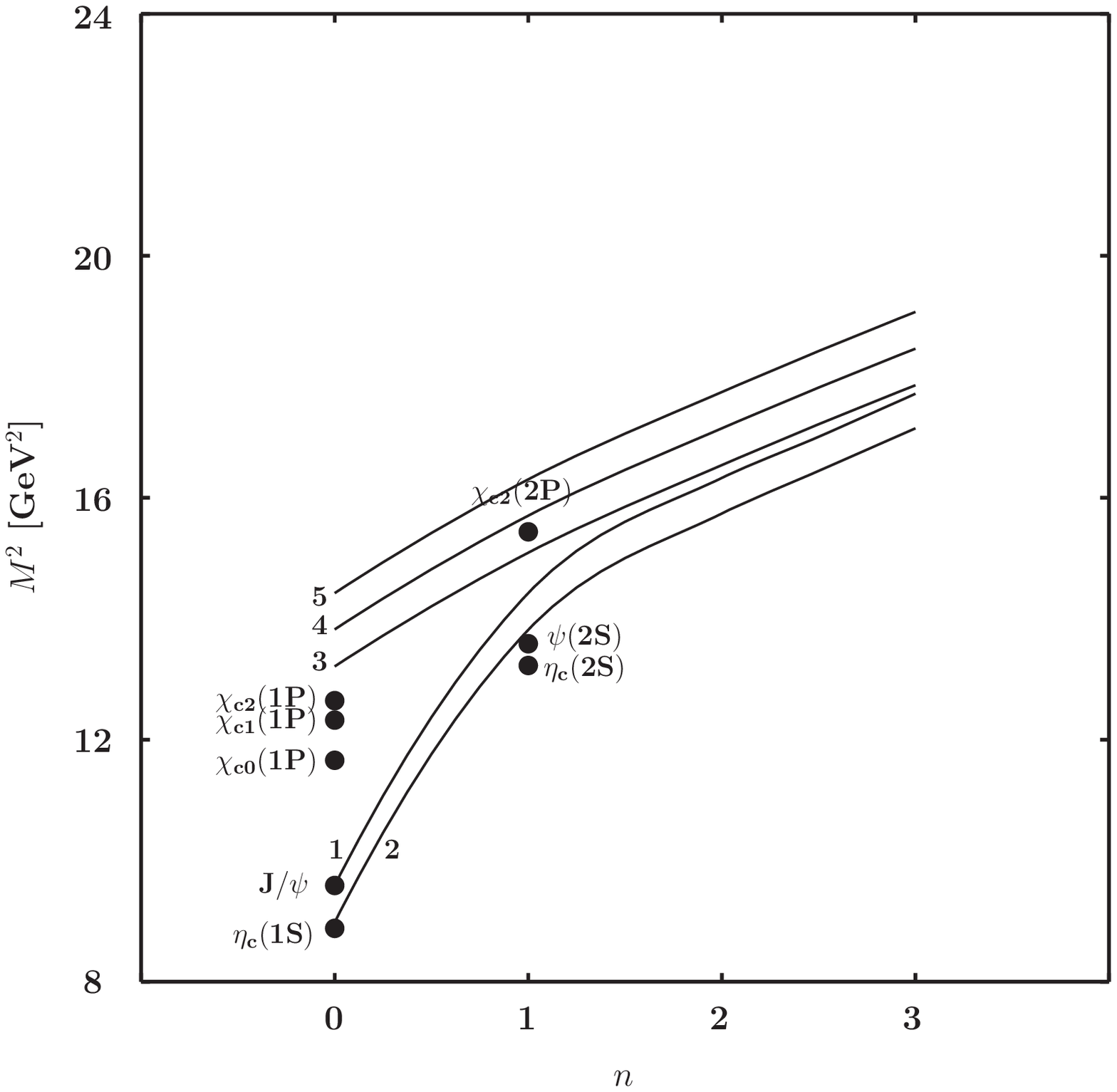}
\caption{Mass spectrum of charmonium states. Central values of data 
are indicated by black circles. 
The numbers mark the corresponding family trajectories: 
1 - $\psi$ mesons; 
2 - $\eta_c$ mesons; 
3 - $\eta_{c0}$ mesons;
4 - $\eta_{c1}$ mesons;
5 - $\eta_{c2}$ mesons. 
}
\label{fig5}

\vspace*{.75cm}

\includegraphics[scale=0.6,angle=0]{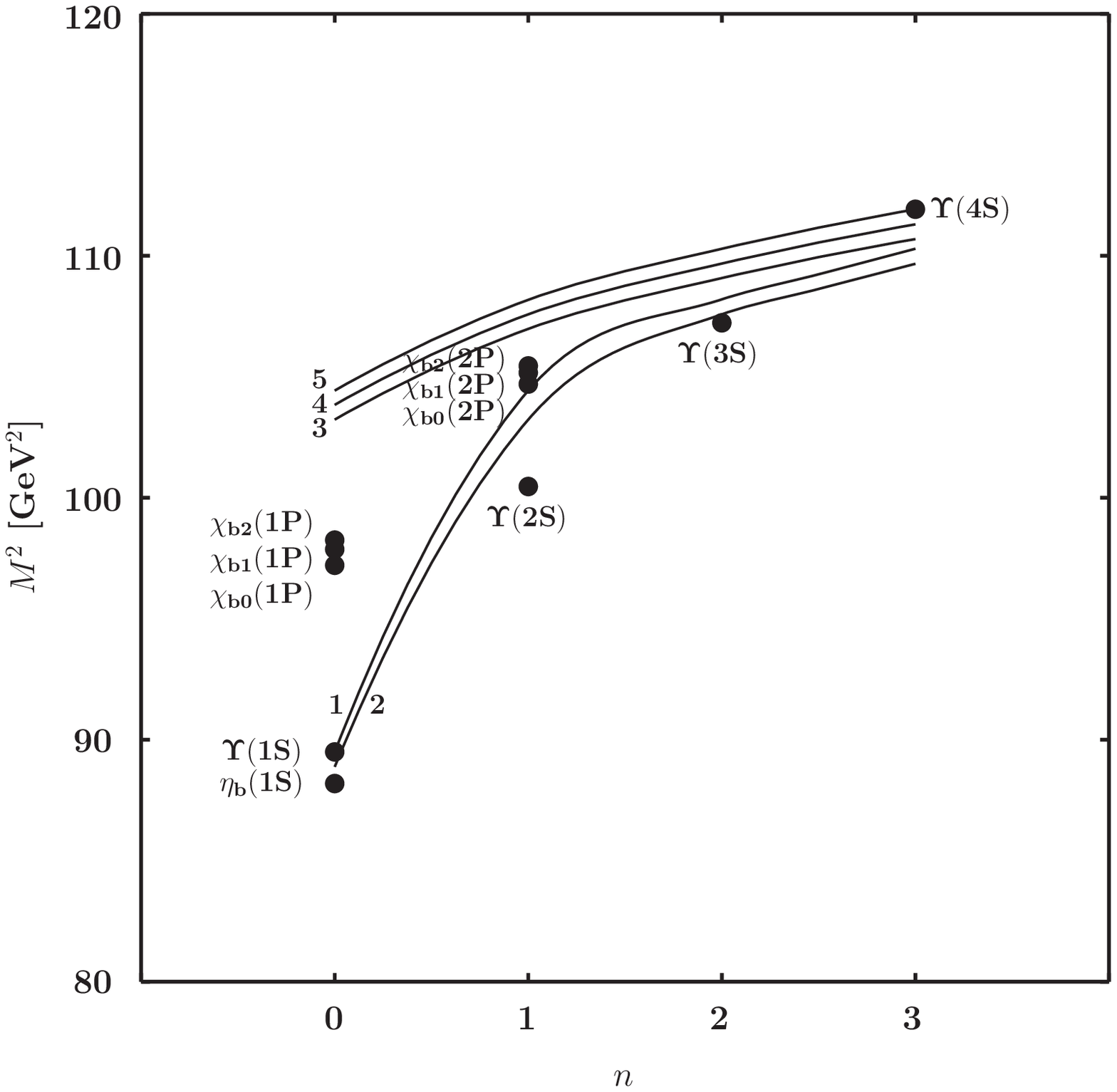}
\caption{Mass spectrum of bottomium states. 
Central values of data are indicated by black circles. 
The numbers mark the corresponding family trajectories: 
1 - $\Upsilon$ mesons; 
2 - $\eta_b$ mesons; 
3 - $\chi_{b0}$ mesons;
4 - $\chi_{b1}$ mesons;
5 - $\chi_{b2}$ mesons. 
} 
\label{fig6}
\end{figure}

\end{document}